# Origin of destruction of multiferroicity in $Tb_2BaNiO_5$ by Sr doping and its implications

Ram Kumar[1], S. Rajput[2], T. Maitra[2], A. Hoser[3], S. Rayaprol[4], Sanjay K Upadhyay[1], K. K. Iyer[1], K. Maiti[1] and E. V. Sampathkumaran[1,4,5,*]

[1]*Tata Institute of Fundamental Research, Homi Bhabha Road, Colaba, Mumbai 400005, India*
[2]*Department of Physics, Indian Institute of Technology, Roorkee-247667, Uttarakhand, India*
[3]*Helmholtz-Zentrum Berlin für Materialien und Energie, Berlin, D-14109, Germany*
[4]*UGC-DAE-Consortium for Scientific Research, Mumbai Centre, BARC Campus, Trombay, Mumbai 400085, India*
[5]*Homi Bhabha Centre for Science Education, Tata Institute of Fundamental Research, V. N. Purav Marg, Mankhurd, Mumbai, 400088 India*

*Corresponding author: *sampathev@gmail.com*

*Abstract*

The orthorhombic Haldane-spin chain compound $Tb_2BaNiO_5$ (Néel order, $T_{N1}$= 63 K) has been shown to be an exotic multiferroic system below ($T_{N2}$=) 25 K due to various fascinating features, pointing to a strong potential for the advancement of concepts in this field. In particular, the rare-earth ions play a direct decisive role unlike in many other well-known multiferroic materials and there appears to be a critical canting angle, developing below $T_{N2}$, subtended by Tb 4f and Ni 3d moments to trigger this cross-coupling phenomenon. However, for a small replacement of Sr for Ba, viz. in $Tb_2Ba_{0.9}Sr_{0.1}NiO_5$, ferroelectricity was reported to get destroyed, but retaining magnetic features at ($T_{N1}$=) 55 K and ($T_{N2}$ =) 14 K. In this article, we address the origin of suppression of multiferrocity in this Sr-doped system through neutron diffraction studies and density functional theory calculations. We find that, unlike in $Tb_2BaNiO_5$, there is no pronounced change in the relative canting angle of the magnetic moments around $T_{N2}$ and that the absolute value of this parameter down to 2 K fails to exceed the critical value noted for the parent, thereby explaining the origin of destruction of magnetoelectric coupling in the Sr-doped material. This finding renders strong support to the proposal of possible existence of 'critical canting angle' – at least in some cases – to induce multiferroicity, apart from serving as a route to engineer multiferroic materials for applications.

## 1. Introduction

There are considerable efforts in the current literature to identify materials with strong magnetoelectric coupling at room temperature and to find mechanisms and routes to engineer such materials. In this respect, multiferroic and magnetodielectric (MDE) properties in the spin-chain compounds, $R_2BaNiO_5$ (R= Nd, Sm, Gd, Dy, Er, Ho and Tb), originally considered [1] to be prototype for Haldane spin-chain behavior [2], are of great interest, as these have been shown to present a variety of interesting situations due to magnetoelectric cross-coupling [3-7]. These compounds crystallize in the orthorhombic, space group *Immm*) [8, 9]. Among these compounds, the behavior of Tb compound, $Tb_2BaNiO_5$, is found to be most spectacular in its magnetic and MDE coupling behavior [4-7]. This compound exhibits magnetic anomalies at two temperatures ($T_{N1}$ = 63 K and $T_{N2}$ = 25 K), with the onset of antiferromagnetic (AFM) ordering temperature, $T_{N1}$, being the largest within this family, attributable to the dominant role of single-ion 4f anisotropy on magnetism. The magnitude of MDE coupling (18%) following the appearance of electric polarization below $T_{N2}$ [4] is the largest ever reported among polycrystalline multiferroic materials at the time of its discovery (though Co analogue was subsequently reported to show still larger values [10]). Another intriguing observation is that the rare-earth was shown to play a direct decisive role to trigger multiferroicity as demonstrated by gradual depression of multiferroic onset temperature by Y-doping (for Tb) [6], unlike in most other well-known multiferroics in which transition metal ions are responsible for this phenomenon. As emphasized by Dong et al [11], such a discovery offers a scope to understand unusual roles of spin-orbit coupling. While these characterize that this Tb compound is an exotic case for such a cross-coupling phenomenon, the intriguing property relevant to the aim of this article is that collinear 3*d*(Ni) and collinear 4*f*(Tb) moments, though mutually canted at all temperatures below $T_{N1}$, lead to multiferroicity only below another characteristic temperature ($T_{N2}$); *the only change in the magnetic structure at $T_{N2}$ appears to be that the canting angles (θ) with respect to c-axis as well as relative canting angle (Δθ) of Tb and Ni moments undergo a dramatic change with the latter showing a drastic increase below $T_{N2}$.* It is difficult to explain (see Ref. 5 for details) such a critical (relative) canting angle induced multiferroicity within conventional Dzyaloshinski-Moriya interaction (DMI) based models [12]; this situation demands further development of theories in the field of magnetoelectric coupling. Considering such a far-reaching consequence for the current understanding of the mechanisms of multiferroicity, we consider it important to provide as much evidence as possible in support of the proposal of critical canting angle for the emergence of ferroelectricity. In this respect, it is important to note that systematic studies [7] of $Tb_2Ba_{1-x}Sr_xNiO_5$ solid solution down to 2 K revealed disappearance of spontaneous electric polarization for a small replacement of Ba by Sr, say to the tune of 10 atomic percent; however, magnetic anomalies at lowered temperatures are noted with corresponding $T_{N1}$ and $T_{N2}$ being 55 and 14 K respectively. [Incidentally, further studies of $Tb_2Ba_{1-x}Sr_xNiO_5$ solid solution suggested [13] that the *x* value where ferroelectricity disappears can be a bit sample dependent, extending up to 0.15]. Thus, this Sr-doped composition (*x*= 0.1) offers an opportunity to address the above issue. This prompted us to carry out detailed neutron powder diffraction (NPD) experiments as a function of temperature (*T*) down to 1.8 K and to augment the conclusions by carrying out density functional theory (DFT) calculations. We find that, while the overall magnetic structure is similar to that of the parent compound, the canting angles of sublattices behave differently in the sense that there is no dramatic change across $T_{N2}$ and that the relative canting angle is not well beyond the critical value [5]. This finding serves as a further support for the proposal on the existence of critical canting angle mooted recently.

## 2. Experimental details

Polycrystalline specimen of $Tb_2Ba_{0.9}Sr_{0.1}NiO_5$ for neutron diffraction measurements was prepared by a standard solid-state reaction method as described earlier [4]. Stoichiometric amounts of $Tb_2(CO_3)_2 \cdot nH_2O$ (99.9%), NiO (99.995%), $BaCO_3$ (99.997%) and $SrCO_3$ (99.9%) have been used as starting precursors. X-ray diffraction (XRD) (Cu-Kα) measurements were performed to confirm the phase formation and crystalline symmetry (*Immm*) of the sample which are in agreement with our previous studies. For this purpose, the room temperature powder x-ray diffraction was subjected to both LeBail and Rietveld analysis methods. The specimen was further characterized by *dc* magnetization measured as a function of both magnetic-field and temperature to confirm that the features reported earlier [7] are reproduced in this freshly prepared specimen. NPD patterns were obtained using the wavelength, λ=2.451 Å, at several temperatures on E6 diffractometer at Helmholtz-Zentrum Berlin (HZB). About 4 g of powder was packed in a vanadium can and attached to a sample holder insert, which was then loaded into a standard liquid helium orange cryostat for obtaining temperature dependent neutron diffraction patterns. The data for each spectrum was collected in 48 scan steps of the 2 two-dimensional area detectors. The total angular range covered (in 2θ) is from 4º to 136º [14]. The specimen was initially cooled to 1.8 K and the diffraction patterns were recorded at selected temperatures during warming cycle. The NPD data was refined for crystallographic as well as magnetic structures using Rietveld refinement method as captured in FullProf suite programs [15]. First-principles DFT calculations were performed to look at the magnetic ground state of the doped compound using the projector augmented plane-wave (PAW) based method as implemented in the Vienna Ab initio Simulation Program (VASP) [16].

## 3. Results and discussion
### 3.1. Neutron diffraction

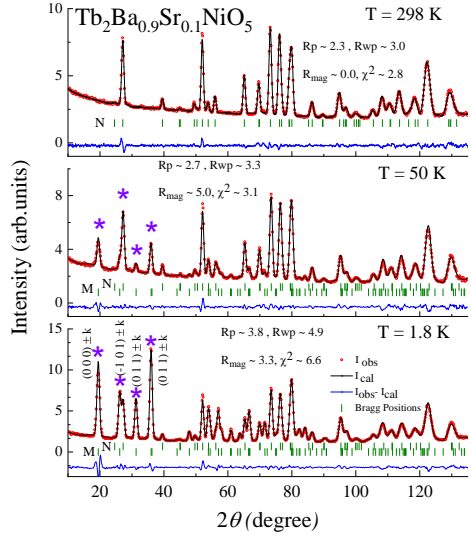

FIG. 1. Rietveld refinement of neutron-diffraction patterns of $Tb_2Ba_{0.9}Sr_{0.1}NiO_5$ at selected temperatures. In the bottom panel, the (*hkl*) values for magnetic Bragg peaks with $k$ = (½, 0, ½) are shown. Blue asterisks mark magnetic peaks. Nuclear (N) and magnetic (M) peak positions are shown by vertical green ticks. The usual parameters obtained by fitting are inserted in the figures.

Before we present the results of neutron diffraction studies, we would like to state the following. In order to verify whether there is any subtle change in crystalline symmetry due to magnetic ordering, we carried out synchrotron powder x-ray diffraction (S-PXRD) measurements down to 2K using x-ray radiation with wavelength 0.883Å on Indian beamline, BL-18B at High Energy Accelerator Research Organization (KEK)-Photon Factory Japan. The beamline energy used was, $E = 14.02$ keV and calibrated using $LaB_6$ standard sample data. We could not resolve any change in the symmetry below $T_{N1}$ or $T_{N2}$. Since there is no other worthwhile feature in the data, we do not present these in this article.

Thus, after eliminating the role of any crystallographic structural change to the loss of ferroelectricity in $Tb_2Ba_{0.9}Sr_{0.1}NiO_5$, we now focus our attention on the magnetic structure and the the canting angles of the magnetic moments of rare earth (Tb 4$f$) and transition metal ions (Ni 3$d$) with respect to $c$-axis, i.e., $\theta$(Tb) and $\theta$(Ni), respectively, and the difference between them $|\Delta\theta|$. The structural phase for the NPD data was refined by Rietveld method using the same structural model used for PXRD refinement. Magnetic structure was refined using the model used for $Tb_2BaNiO_5$ [5].

In Fig. 1, the observed ($I_{obs}$) NPD patterns along with refined profile ($I_{cal}$) are shown for $T$ = 1.8, 50, and 298 K. The figures are shown along with the difference pattern and the Bragg peak positions allowed by the space group symmetry. The NPD pattern collected at $T$ = 298 K was refined successfully for the orthorhombic crystal structure in the space group $Immm$, and is in good agreement with the previous reports [5, 9].

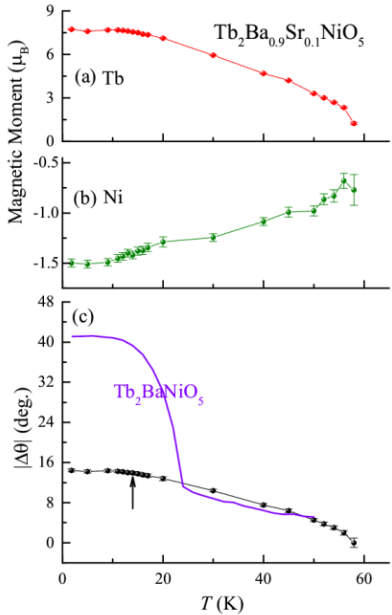

FIG. 2. **(a), (b)** Temperature dependence of Tb and Ni moments respectively and the minus sign for Ni is placed to indicate that it is oriented opposite to Tb moments. **(c)** Temperature dependence of relative canting angle of Ni and Tb in $Tb_2Ba_{0.9}Sr_{0.1}NiO_5$ obtained from NPD data, with the vertical arrow denoting $T_{N2}$; the profile of the plot for the parent compound reported in Ref. 5 is also shown for the sake of comparison to bring out how the relative canting angle jumps at its respective $T_{N2}$.

At $T = 50$ K, the structural phase can be refined using the 298 K model. However, there are a few extra Bragg peaks appearing in the diffraction pattern, all of which could be accounted for by invoking the magnetic structural phase with the propagation vector, $k = (½, 0, ½)$. It may be noted here that there is no signature of any change in the crystallographic structure or of a reduction in the symmetry as a function of $T$ and the crystallographic Bragg peaks at the lowest temperature are the same as those at the highest temperature measured. As the temperature is lowered below $T_{N1} \sim 53$ K, the intensities of magnetic Bragg peaks (with respect to nuclear Bragg peaks) gradually get enhanced due to a steady increase in the ordered moments of Tb and Ni, as shown in Fig. 2a and 2b (derived from the present data). See the peaks marked by (*) in Fig. 1. As the material is cooled below $T_{N2}$, these intensities are found to increase further, however, without any indication for any change in the magnetic structure. Fig. 2(c), shows the variation in the relative magnitude of angle ($\Delta\theta$) for Ni and Tb magnetic moments as a function of temperature.

It is to be noted that Tb moment lies almost along $c$-axis (nearly $< 5°$) as in the case of parent within the limits experimental error (few degrees), whereas the canting angle for Ni moment remains around $22°$. This angle for Ni undergoes a very weak decrease with $T$ unlike for the parent in which case Ni moment moves away from $c$-axis with decreasing temperature (saturating to a value of about $40°$ below about 15 K [5]). The magnetic structures at 1.8, 30, and 50 K are shown in Fig. 3. [The results are in good agreement with the theoretically calculated magnetic structure discussed below including the canting angles (see Fig. 2c for experimental data)]. There is no worthwhile anomaly in the values of the magnetic moments at $T_{N2}$ within the limits of experimental error, except that there is a saturation below this temperature. In any case, the most notable finding is that there is no sharp change in $\theta$ of these moments throughout the temperature range, in particular across $T_{N2}$ (14 K), although these moments are weakly canted in the entire $T$ range below $T_{N1}$. Thus, there is a marked difference in the canting angle behavior with respect to that reported for the parent (see Fig. 2c). The net result is that $|\Delta\theta|$ increases only gradually with the lowering of temperature, attaining a value of about $14°$ at $T_{N2}$ and saturating thereafter for this Sr-doped composition. This value, though very close to the critical canting angle noted for the parent compound ($>15°$), is barely enough to develop distinct spontaneous electric polarization. [But, the proximity to critical angle could be the reason why we observed a small spread ($x= 0.1$ to $0.15$) in the composition at which ferroelectricity can be destroyed]. This is the most crucial finding to the aim of this article and establishes that the canting angle subtended by Ni and Tb with $c$-axis is very critical for the magnetoelectric coupling / multiferroicity.

### *3.2. DFT calculations*

We now present the results of first-principles DFT calculations to infer the magnetic ground state of the doped material. We considered a doping of 0.125 of Sr at Ba sites (which is close to the experimental doping (of 0.1) reported in this article) in order to minimize the computational cost. To carry out the calculations, optimization of experimental structure has to be done as the system is doped. So, structural optimization of atomic coordinates (keeping the lattice parameters same as those at their respective experimental values) are carried out until the Hellmann Feynman force on each atom is less than 5 meV /Å and Tb *4f* electrons were considered as core electrons. The self-consistent total-energy calculations on the optimized structure thus obtained were then performed for various magnetic configurations until an energy convergence of $10^{-4}$ eV was achieved. In all the calculations, a $\Gamma$ centered 4x4x4 Monkhorst-Pack **k**-mesh was used for performing the Brillouin zone integrations and an energy cut off of 500eV was used for plane waves. Perdew-Burke-Ernzerhof Generalized Gradient Approximation (PBE-GGA) [17] was considered for exchange-correlation functional and Coulomb correlation (U) and spin-orbit interaction (SO) were considered within GGA+$U$+SO approximations as implemented in VASP [16]. We computed total energies for collinear (AFM-CL) configuration as well as for non-collinear (AFM-NCL) spin configuration (where Ni spins are canted with respect to the $c$ axis in the a-c plane resembling to that seen experimentally at 1.8 K, see Fig. 3) for the AFM order within the GGA+$U$+SO approximation for various U values.

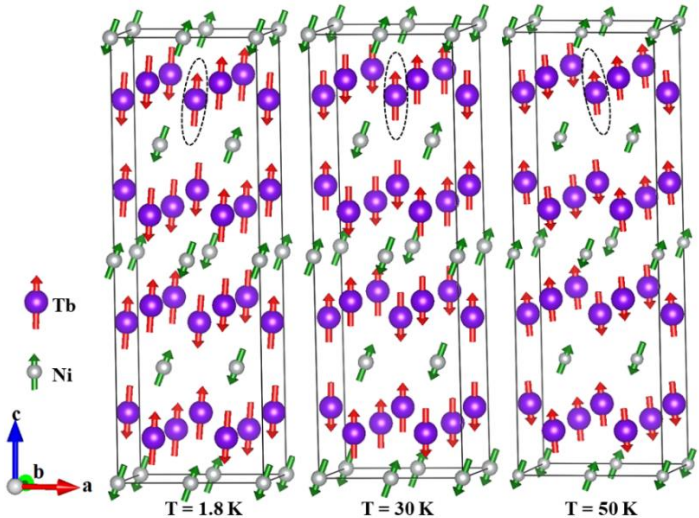

FIG. 3. Magnetic structure of $Tb_2Ba_{0.9}Sr_{0.1}NiO_5$ at selected temperatures. A 2×1×2 supercell is shown here. Dotted black circles around a Tb are drawn to show that there is a marginal change in the angle with respect to basal plane with varying temperature, with the result that the moment is almost along $c$-direction.

The magnetic moments of both Tb and Ni ions were fixed along $c$-axis for AFM-CL magnetic state. On the other hand, for AFM-NCL magnetic state, the Ni moment was allowed to relax in the $a$-$c$ plane over a self-consistency cycle until it reached the minimum energy value, while the Tb moment was kept fixed along the $c$-axis. For Tb 4$f$ states, we have used $U_{eff}$ = U – J = 7eV where U is the Coulomb correlation and J is Hund's exchange as was previously used for the parent compound $Tb_2BaNiO_5$ [5]. The value of $U_{eff}$ at Ni site for $d$ states is then varied in the range 0 to 2.5 eV for AFM-CL and AFM-NCL magnetic configurations. The relative values of total energies between AFM-CL and AFM-NCL are listed in Table I for various $U_{eff}$ values at Ni site.

**Table I.** Relative energy (in meV) of AFM-CL and AFM-NCL configurations obtained within GGA+U+SO approximation for different $U_{eff}$ values for Ni d states while keeping $U_{eff}$ = 7eV fixed for Tb 4f states.

| $U_{eff}$ (eV) | Tb = 7.0, Ni = 0.0 | Tb = 7.0, Ni = 1.5 | Tb = 7.0, Ni = 2.5 |
|---|---|---|---|
| AFM-CL | 0.0 | 84.42 | 0.0 |
| AFM-NCL | 0.18 | 0.0 | 43.62 |

Comparing total energies of AFM-CL and AFM-NCL magnetic states, we observe that the two magnetic states are almost degenerate when $U_{eff}$ = 0 at Ni site. Increasing $U_{eff}$ to 1.5 eV stabilizes the AFM-NCL magnetic state. However, at $U_{eff}$ = 2.5 eV the ground state changes to AFM-CL state similar to that observed in case of the parent compound [5]. Comparing the relative energy differences between AFM-NCL and AFM-CL for the Sr doped compound with those for parent compound, we see that the energy differences in Sr doped compound are smaller than the corresponding values for the parent compound which goes well with the lesser magnetic transition temperatures seen in the doped case.

TABLE II. Canting angle of $Ni^{2+}$ for $Tb_2BaNiO_5$ and $Tb_2Ba_{0.875}Sr_{0.125}NiO_5$ within GGA+U+SO as a function of $U_{eff}$ applied to Ni ($U_{eff}$ for Tb is kept fixed at 7eV). Canting angle of $Ni^{2+}$ is decreased with Sr doping. The value of angle for $U$=1.5 eV is close to the experimentally observed value.

| $U_{eff}$ (eV) | $\theta$ (degree) | |
|---|---|---|
| | $Tb_2BaNiO_5$ | $Tb_2Ba_{0.875}Sr_{0.125}NiO_5$ |
| 0.0 | 20.28 | 14.52 |
| 1.5 | 30.90 | 13.71 |
| 2.5 | 38.07 | 28.81 |

We have also computed $\theta$ of Ni spin moments with respect to *c*-axis for our AFM-NCL calculations which is seen to change with $U_{eff}$ applied to Ni (See Table II). By comparing the canting angles of Ni spin in parent and doped compound, we clearly observe that the canting angles remain consistently much lower in the case of Sr-doped material for all the $U_{eff}$ values studied. This is consistent with the experimental findings discussed above. For $U_{eff}$ = 1.5 eV, where AFM-NCL is the ground state, the canting angle is found to be 13.71° which is close to experimentally observed value. We can conclude from our study that such a small substitution of Sr at Ba site brings down the relative canting angle of Tb and Ni below the critical canting angle required for multiferroicity while keeping the intra-sublattice collinear magnetic ground state unaltered.

### 4. Summary

A combined look at the physical properties observed experimentally and theoretical calculations on $Tb_2Ba_{0.9}Sr_{0.1}NiO_5$ and a comparison with those of $Tb_2BaNiO_5$, clearly establish that there is a critical canting angle between the magnetic moments of Tb 4f and Ni 3*d* needed to induce electric polarization and therefore multiferroicity in this Haldane spin-chain compound. Besides, this work asserts a possible route to engineer multiferroicity for applications under favorable circumstances in other materials.

**Acknowledgements**
Authors acknowledges financial support from the Department of Atomic Energy, Govt. of India (Grant number 12-R&D-TFR-5.10-0100). SR and KKI thank the Department of Science and Technology, India for the financial support, and Saha Institute of Nuclear Physics and Jawaharlal Nehru Centre for Advanced Scientific Research for facilitating the experiments at the Indian Beamline, Photon Factory, KEK, Japan, under the proposal number JNC/KEK-JAP/IN-56 for synchrotron XRD measurements on Indian beamline (BL-18B) at PF, KEK, Japan. Authors also thank Gouranga Manna and Sabyasachi Karmakar for their help during the Synchrotron X-ray powder diffraction experiments. S. Rajput thanks MHRD (India) for research fellowship. EVS thanks Science and Engineering Research Council (India) for J C Bose Fellowship (Grant number: SR/S2/JCB-23/2007).